\documentstyle[12pt,psfig]{article}
\begin{document}
\begin{center}
\Large{\bf Photon Production in Relativistic Heavy Ion Collisions}
\vskip 0.2in

\large{Dinesh Kumar Srivastava }
\vskip 0.2in

\large{\em Variable Energy Cyclotron Centre,\\
 1/AF Bidhan Nagar, Calcutta 
700 064, India\\
and\\
Fakult\"{a}t f\"{u}r Physik, Universit\"{a}t Bielefeld, D-33501,
Bielefeld, Germany}

\vskip 0.2in

Abstract

\vskip 0.2in

\end{center}

The production of single photons in relativistic heavy ion collisions
at CERN SPS, BNL RHIC and CERN LHC energies is re-examined in view
of the recent studies of Aurenche et al which show that
the rate of photon production from quark gluon plasma, evaluated
at the order of two loops far exceeds the rates evaluated at one-loop level
which have formed the basis of all the estimates of photons so far.
We find that the production of photons from quark matter could easily
out-shine those from the hadronic matter in certain ideal conditions.

\newpage

Single photons can be counted among the first signatures~\cite{first}
  which were proposed to verify the formation of
deconfined strongly interacting matter- namely the quark gluon plasma
(QGP). Along with dileptons- which will have similar origins, they
constitute electro-magnetic probes which are believed to reveal the
history of evolution of the plasma, through a (likely) mixed phase
and the hadronic phase, as they do not re-scatter once produced
and their production cross section is a strongly increasing 
function of temperature. During the QGP phase, the single photons
are believed to originate from Compton 
($q\,(\overline{q})\,g\,\rightarrow\,q\,(\overline{q})\,\gamma$)
and annihilation ($q\,\overline{q}\,\rightarrow\,g\,\gamma$)
processes~\cite{joe,rolf} as well as bremsstrahlung processes
($q\,q\,(g)\,\rightarrow\,q\,q\,(g)\,\gamma$). Recently in the
first evaluation of single photons within a parton cascade model~\cite{pcm},
it was shown~\cite{pcmphot} that the fragmentation of time-like 
quarks ($q\,\rightarrow\,q\,\gamma$) produced
in (semi)hard multiple scatterings during the pre-equilibrium phase 
of the collision  leads to a substantial production of photons (flash of
photons!), whose $p_T$ is decided by the $Q^2$ of the scatterings and not the
temperature as in the above mentioned calculations.

The upper limit for production of single photons in $S+Au$ collisions 
at SPS energies~\cite{wa80} has been used by several authors to rule out simple
hadronic equations of states~\cite{prl} and the final results for the $Pb+Pb$
collisions at SPS energies are eagerly awaited. 

In a significant development Aurenche et al~\cite{pat} have recently
evaluated the production of photons in a QGP up to two loops and shown
that the bremsstrahlung process gives a contribution which is
similar in magnitude to the Compton and annihilation contributions evaluated 
up to the order of one loop earlier~\cite{joe,rolf}. This
is in contrast to the `expectations' that the bremsstrahlung contributions
drop rapidly with energy (see Ref.~\cite{kryz,dipali} for estimates
within a soft photon approximation). They also reported an entirely new
mechanism for the production of hard photons through the annihilation
of an off-mass shell quark and an anti-quark, where the off-mass shell
quark is a product of scattering with another quark or gluon and
which completely dominates the emission of hard photons. 
This process is similar to the annihilation of quarks in the presence of
the chromo-electric field which may develop when two nuclei pass through
each other due to colour exchange, and which can absorb the unbalanced
energy and momentum to ensure the feasibility of the process which is
absent in vacuum~\cite{avijit}.

 If confirmed, this has far reaching consequences
for the search of single photons from the relativistic heavy ion
collisions.

The rate for the production of hard photons evaluated to one
loop order using the effective theory based on resummation of
hard thermal loops is given by~\cite{joe,rolf}:
\begin{equation}
E\frac{dN}{d^4x\,d^3k}=\frac{1}{2\pi^2}\,\alpha\alpha_s\,
                          \left(\sum_f e_f^2\right)\, T^2\,
                       e^{-E/T}\,\ln(\frac {cE}{\alpha_s T})
\end{equation}
where the constant $c\approx$ 0.23.  The summation runs over the
the flavours of the quarks and $e_f$ is the electric charge of the
quarks in units of charge of the electron. The rate of production
of photons due to the bremsstrahlung processes evaluated by
Aurenche et al is given by:
\begin{equation}
E\frac{dN}{d^4x\,d^3k}=\frac{8}{\pi^5}\,\alpha\alpha_s\,
                          \left(\sum_f e_f^2\right)\, 
                        \frac{T^4}{E^2}\,
                       e^{-E/T}\,(J_T-J_L)\,I(E,T)
\end{equation}
where $J_T\approx$ 4.45 and $J_L\approx - $4.26 for 2 flavours and 3
colour of quarks. For 3 flavour of quarks, $J_T\approx$ 4.80 and
 $J_L\approx - $4.52.  $I(E,T)$ stands for;
\begin{eqnarray}
I(E,T)&=&\left[ 3\zeta(3)+\frac{\pi^2}{6}\frac{E}{T}+
        \left(\frac{E}{T}\right)^2\ln(2)\right.\nonumber\\
        & &+4\,Li_3(-e^{-|E|/T})+2\,Li_2(-e^{-|E|/T})\nonumber\\
        & &\left. -\left(\frac{E}{T}\right)^2\,\ln(1+e^{-|E|/T})\right]~,
\end{eqnarray}
and the poly-logarith functions $Li$ are given by;
\begin{equation}
Li_a(z)=\sum_{n=1}^{+\infty}\frac{z^n}{n^a}~~.
\end{equation}

 And finally the contribution of the $q\overline{q}$ annihilation
with scattering obtained by them is given by:
\begin{equation}
E\frac{dN}{d^4x\,d^3k}=\frac{8}{3\pi^5}\,\alpha\alpha_s\,
                          \left(\sum_f e_f^2\right)\, ET \,
                          e^{-E/T}\,(J_T-J_L)
\end{equation}
 
We plot these rates of emission of photons
from a QGP at $T=$ 250 MeV (Fig.~1) for an easy comparison.
The dashed curve gives the contribution of the Compton
and annihilation processes evaluated to the order of one loop by
Kapusta et al~\cite{joe}, the dot-dashed curve gives the bremsstrahlung
contribution evaluated to two-loops by Aurenche et al~\cite{pat} while the
solid curve gives the results for the annihilation with scattering 
evaluated by the same authors. The dotted curve gives the results for
the bremsstrahlung contribution evaluated within a soft-photon
approximation (and using thermal mass for quarks and gluons) obtained
by Pal et al~\cite{dipali}. We see that at larger energies the
annihilation of quarks with scattering really dominates over the
rest of the contributions by more than a order of magnitude.

How much of this dominance does survive when we integrate the radiation of
photons over the history of evolution of the system, specially as the
QGP phase occurring in the early stages of the evolution necessarily
occupies smaller four-volume compared the hadronic matter,
which is known to have an emission rate similar to the quark matter
at a given temperature~\cite{joe} at least when only the Compton and the
annihilation terms are used?

In order to ascertain this we consider central collision of lead nuclei
at SPS, RHIC and LHC energies. We assume that a chemically and
thermally equilibrated quark-gluon plasma is formed at $\tau_0=$ 1
fm/$c$ at SPS and at 0.5 fm/$c$ at RHIC and LHC energies. While there are
indications that the plasma produced at the energies under
consideration may indeed attain thermal equilibrium at around
$\tau_0$ chosen here~\cite{pcm,therm}, it is not quite definite
that it may be chemically equilibrated. It may be recalled 
that the parton cascade model which properly accounts for multiple
scatterings uses a cut-off in momentum transfer and virtuality to
regulate the divergences in the scattering  and the branching amplitudes for 
partons. This could underestimate the extent of chemical equilibration,
 by a cessation of interactions when the energy of the partons is still large
which would not be the case if the screening of the partonic 
interactions could be accounted for.
The self-screened parton cascade~\cite{sspc}
on the other hand attempts to remove these cut-offs by
estimating the screening offered by the partons which have larger
$p_T$ (and hence materialize earlier) to the partons which have
smaller $p_T$ (and hence materialize later). However it does not
explicitly account for multiple scattering except for what is contained 
in the Glauber approximation utilized there. 

In these exploratory calculations we assume a chemical equilibration
at the time $\tau_0$ such that the initial temperature is obtained
from the Bjorken condition~\cite{bj};
\begin{equation}
\frac{2\pi^4}{45\zeta(3)}\,\frac{1}{\pi R_T^2}\frac{dN}{dy}=4 a
T_0^3\tau_0
\end{equation}
where we have chosen the particle rapidity densities as 825, 1734,
and 5625 respectively at SPS, RHIC, and LHC energies for central
collision of lead nuclei~\cite{kms} and taken $a=47.5\pi^2/90$ for
a plasma of mass-less quarks (u, d, and s) and gluons.

We assume the phase transition to take place at $T=$ 160 MeV, and the
freeze-out to take place at 100 MeV.
We use a hadronic equation of state consisting of all the hadrons and
resonances from the particle data table which have a mass less then 2.5
GeV~\cite{jean}. The rates for the hadronic matter have been 
obtained~\cite{joe}
from a two loop approximation of the photon self energy 
using a model where $\pi-\rho$ interactions have been included. The 
contribution of the $A_1$ resonance is also included according to the
suggestions of Xiong et al~\cite{li}. The relevant hydrodynamic equations are
solved using the procedure~\cite{hydro} discussed earlier and
a integration over history of evolution is performed~\cite{jean}. 

In Fig.~2 we show our results for central collision of lead nuclei
at energies which are reached at CERN SPS. We give the
contribution of the quark
matter (from the QGP phase and the mixed phase) labeled as QM
and that of the hadronic matter (from the mixed phase and the hadronic
phase) separately.
 We see that if we use the rates obtained earlier by Kapusta et
al, there is no window when the radiations from the quark-matter could
shine above the contributions from the hadronic matter.  However, once the
newly obtained rates are used we see that the quark matter may 
indeed out-shine the hadronic matter up to $p_T=$ 2 GeV, from these
contributions alone.  
Note that by tracking the history from $\tau_0$= 1 fm/$c$ onward, we
have not included the pre-equilibrium contributions~\cite{pcmphot} which
will make a large contribution at higher momenta. 
The contribution of hard QCD photons~\cite{qcd} obtained  by scaling 
the results for $pp$ collisions by the nuclear thickness.

The results for RHIC energies (Fig.~3) are quite interesting as now the window 
over which the quark matter out-shines the hadronic contributions
stretches to almost 3 GeV. Once again the addition of the pre-equilibrium
contributions at larger $p_T$ would substantially widen this window.

At LHC energies this window extends to beyond 4 GeV, and considering
that perhaps the local thermalization at LHC (and also at RHIC) could
be attained earlier than what is definitely a very conservative value
here, these results provide the exciting possibility that if these
conditions are met the quark matter may emit photons which may be
almost an order of magnitude larger than those coming from the
hadronic matter over a fairly wide window. As mentioned earlier, the
pre-equilibrium contribution (due to the very larger initial energy)
should be much larger here and we may have the exciting possibility that
the quark matter may out-shine the hadronic matter over a very large
window indeed.

How will the results change if the QGP is not in chemical equilibrium?
While it is not easy to perform the estimates similar to the one 
done by Aurenche et al for a chemically non-equilibrated plasma,
it is reasonable to assume that the rates will fall simply because
then the number of quarks and gluons will be smaller. Some of this
short-fall will be off-set by the much larger temperatures 
which the parton cascade models predict. If one considers
a chemically equilibrating plasma~\cite{smm} then the quark and
gluon fugacities will increase with time and at least the
contributions from the latter stages will not be strongly suppressed.
It is still felt that the loss of production of high $p_T$ (from
early times) photons due to chemical non-equilibration would be
more than off-set by the increased temperature  and the pre-equibrium
contribution, which can be quite large.  

We conclude that the newly obtained rates for emission of
photons from QGP (evaluated to the order of two loops) suggest that
if chemically equilibrated plasma is produced then there will exist
a fairly wide window where the photons from quark matter
may outshine the photons from hadronic matter. Even in the absence
of chemical equilibration these results indicate an enhanced
radiation from the quark matter which is of considerable interest.

\section*{Acknowledgments} 
The author gratefully acknowledges the hospitality of University of
Bielefeld where part of this work was done. He would also like to acknowledge
useful discussions with Jean Cleymans and Francois Gelis. He is
especially grateful to Haitham Zaraket for suggesting
that the exact expression for the bremsstrahlung contribution be used.

\newpage

\begin{figure}
\psfig{file=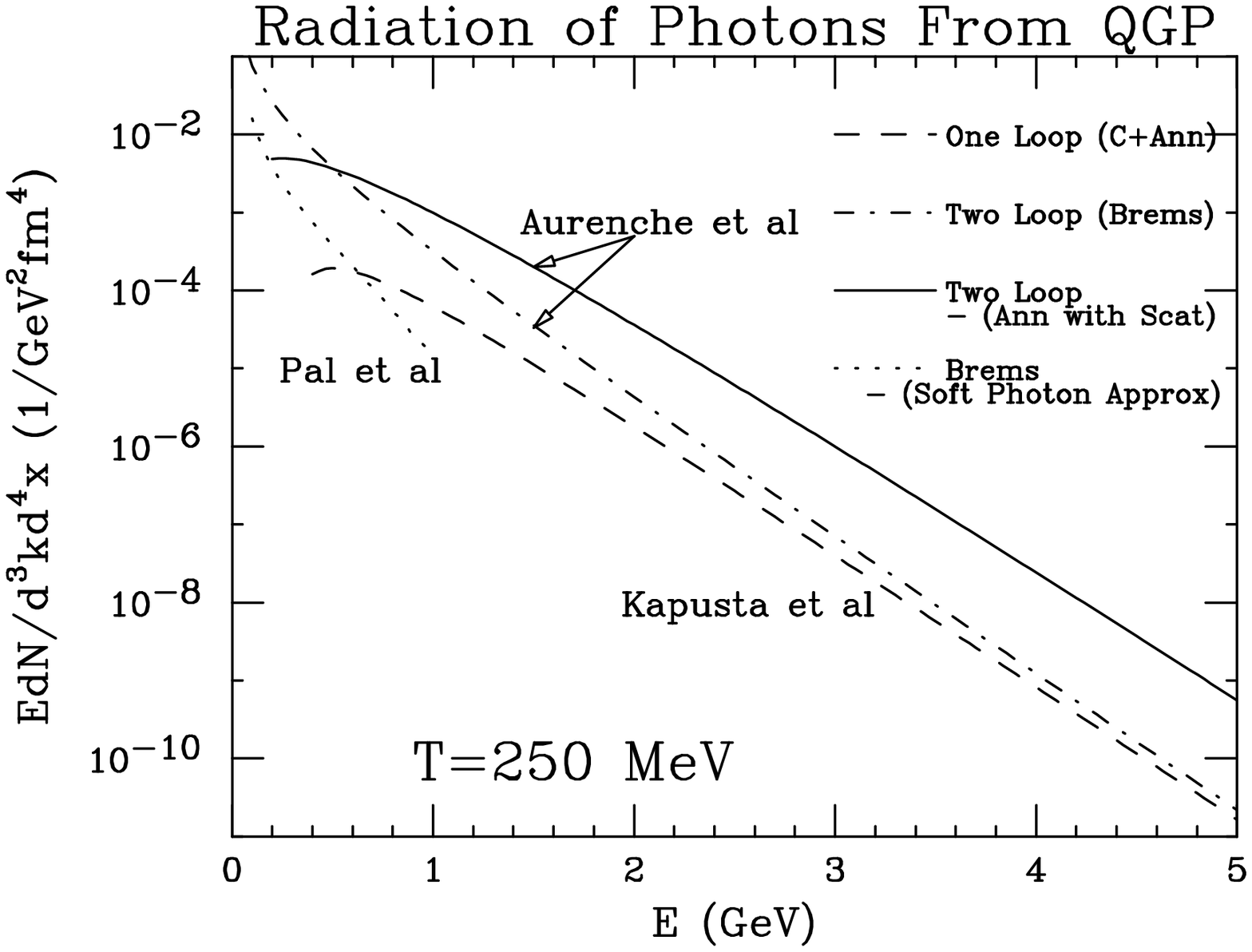,height=12cm,width=15cm}
\vskip 0.1in
\caption{ Radiation of photons from various processes in the quark
matter at $T=$ 250 MeV}
\end{figure}

\newpage

\begin{figure}
\psfig{file=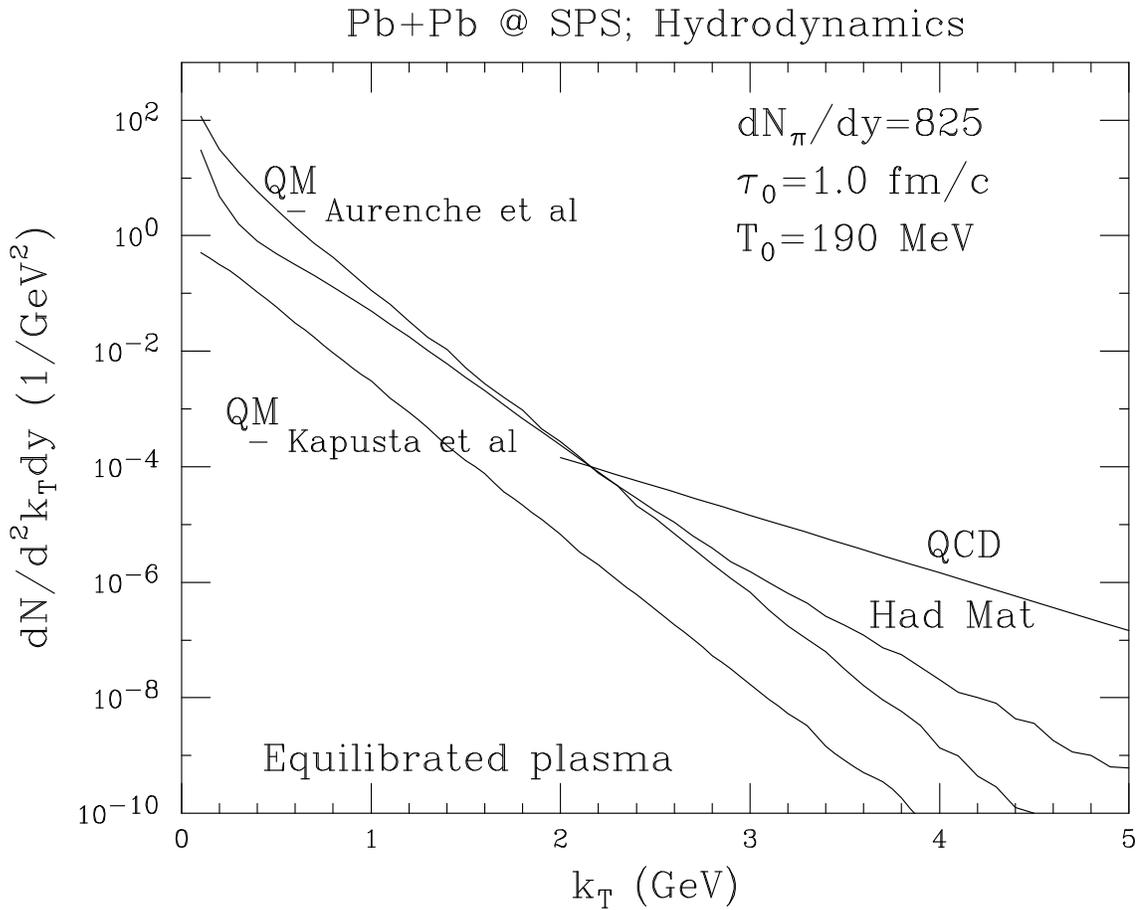,height=12cm,width=15cm}
\vskip 0.1in
\caption{ Radiation of photons from central collision of lead nuclei 
at SPS energies from the hadronic matter (in the mixed phase and the
hadronic phase) and the quark matter (in the QGP phase and the mixed
phase).
The contribution of the quark matter while using the
rates obtained by Kapusta et al and Aurenche et al,
and those from hard QCD processes 
 are shown separately
}
\end{figure}
\newpage
\begin{figure}
\psfig{file=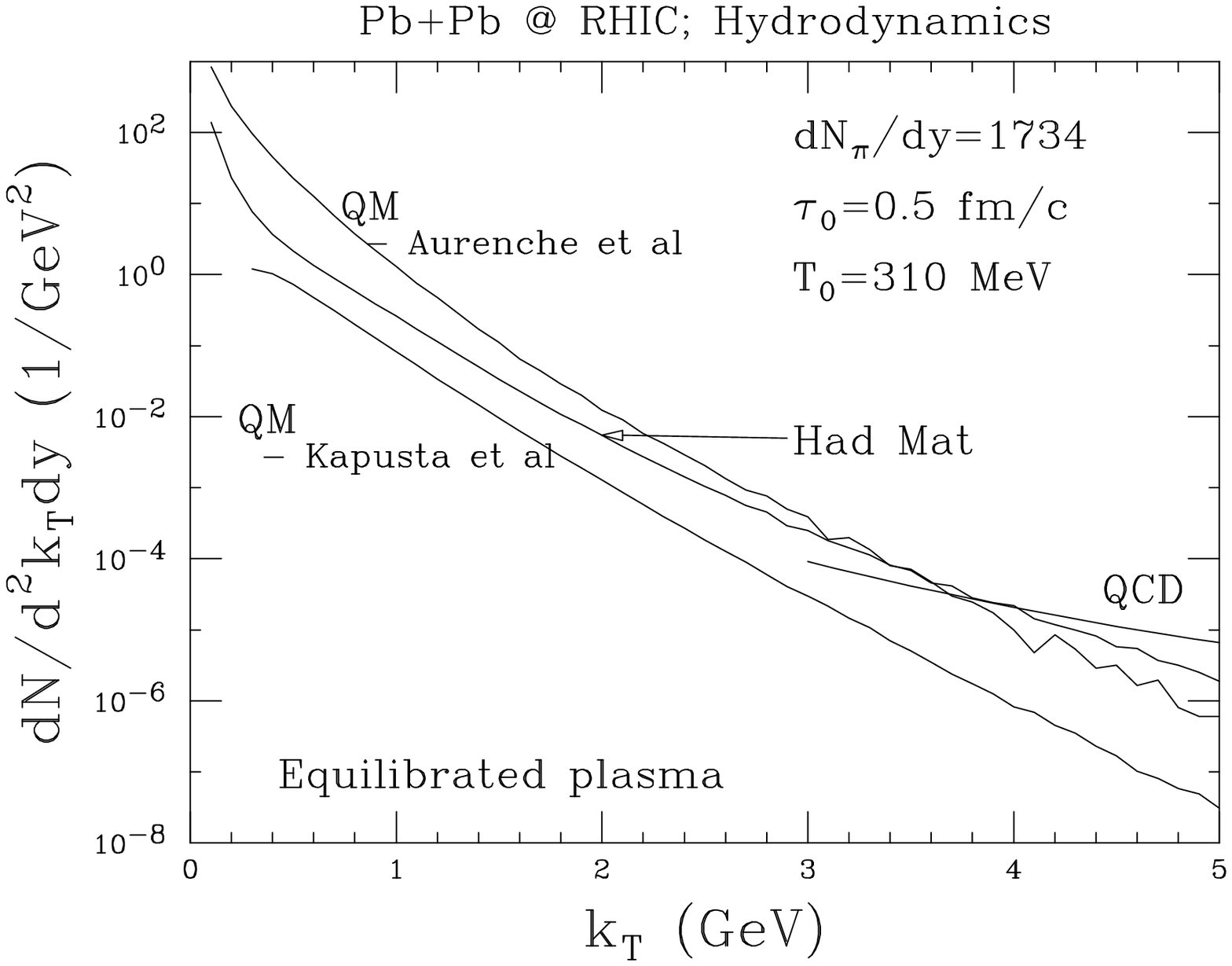,height=12cm,width=15cm}
\vskip 0.1in
\caption{ Same as Fig.~2 for RHIC energies.
}
\end{figure}
\newpage
\begin{figure}
\psfig{file=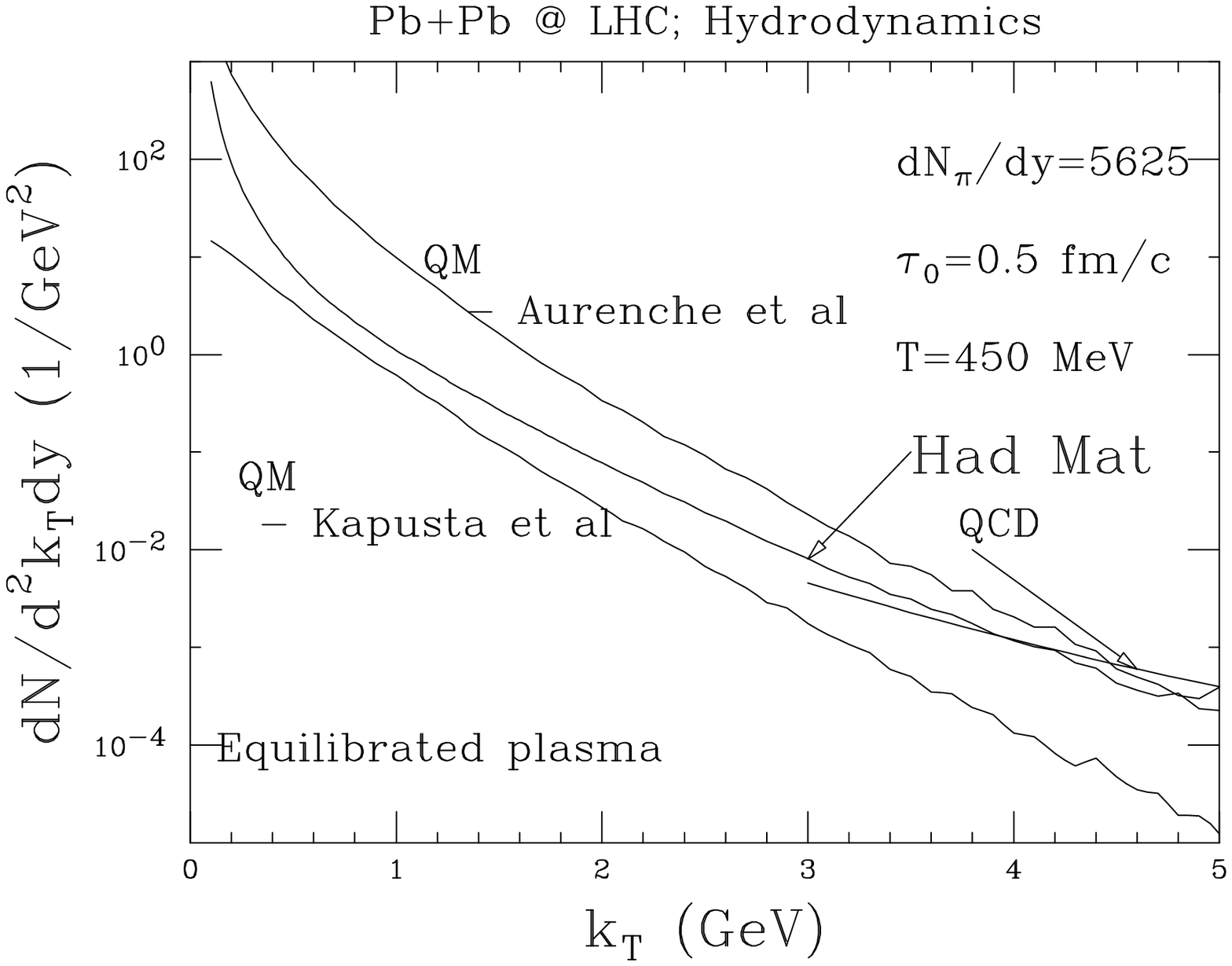,height=12cm,width=15cm}
\vskip 0.1in
\caption{ Same as Fig.~2 for LHC energies. 
}
\end{figure}
\end{document}